\documentstyle[12pt,epsf]{article}
\def\g{\gamma}
\def\a{\Gamma}
\def\pO{\pi^0}
\def\pP{\pi^+}
\def\pM{\pi^-}
\def\gn{\g d \rightarrow \pO +\g pn}

\def\gp{\g d\rightarrow \pM +\g pp}
\def\gnn{\g d\rightarrow \pP +\g nn}

\def\dP{\g d\rightarrow \pM D}
\def\dN{\g d\rightarrow \pP D}
\def\O{d\Omega}
\def\C{d\sigma_{\dP}}
\def\B{d\sigma_{\dN}}
\def\CO{\frac{\C}{\O}}
\def\BO{\frac{\B}{\O}}

\def\l{(\mu b)}
\def\te{\theta_{\pi}}
\def\tg{\theta_{\g}}
\def\tn{\theta_n}

\def\om{\omega}
\def\dn{D\rightarrow \g NN}
\def\be{\begin{equation}}
\def\ee{\end{equation}}
\def\beq{\begin{eqnarray}}
\def\eeq{\end{eqnarray}}

\begin{document}
\pagestyle{empty}
\setcounter{page}{0}           %Engl TITLE
\begin{center}
{\large\bf P.N.Lebedev Physical Institute}\\
\vspace{2.0cm}
{\bf High energy and cosmic rays physics}\\
\vspace{.5cm}
{\bf High energy physical department}\\
\end{center}
\vspace{2.5cm}
\begin{flushright}
    Preprint N~~{\LARGE 52}
\end{flushright}
\vspace{.5cm}
\begin{center}
{\large V.M.~Alekseyev, S.N.~Cherepnya, L.V.~Fil'kov and V.L.~Kashevarov}
\end{center}
\vspace{1.5cm}
\begin{center}
 {\Large \bf Possibility of observation of supernarrow\\
\vspace{0.2cm}
 dibaryons with symmetric wave function\\
\vspace{0.2cm}
 in $\g d\rightarrow \pi^{\pm} + \g NN$ reactions}
\end{center}
\vspace{4.5cm}
\begin{center}
Moscow-1996
\end{center}

\newpage
\large
\vspace{2cm}
\begin{center}
{\bf Abstract}
\end{center}

We study the possibility of observation of supernarrow dibaryons with a
symmetric wave function in reactions $\gnn$ and $\gp$. The method of a
dibaryon masses reconstruction over the dibaryon decay products $(\g nn)$
and $(\g pp)$ is analysed. The Monte-Carlo simulation is used to choose the
optimal location of the detectors, to estimate the contribution of the
main background processes and to calculate the expected yields of the
dibaryon production. It is shown that 100 hours of the exposition time at
the microtron MAMI (Mainz) is quite enough to determine whether or not
the supernarrow dibaryons with symmetric wave function exist.

\normalsize
\newpage
\pagestyle{plain}
\setcounter{page}{3}
\section{Introduction}\
\vspace{1cm}

The possibility of existence of many quarks states is predicted by QCD [1].
This is a new kind of the nuclear matter. Of particular interest are narrow
six-quark states--dibaryons with decay width of less than a few MeV.
The experimental discovery of such states would have important consequences
for particle and nuclear physics. An intensive search for narrow dibaryons
states has been performed and a number of candidates for this states has
been found (see for example [2--12]).
Unfortunately, the observed effects are not big and
a background is big and uncertain. So it is difficult to claim
unequivocally that dibaryon states have actually been found [13].
Therefore, if we want to make unambiguous conclusions about the existance
of dibaryons we must regard such type of dibaryons and such processes
where the contribution of the dibaryons is very big and a background is
small.

We suggest to search dibaryons with symmetric wave function [14--19].
Such dibaryons satisfy the following condition:
\begin{equation}
 (-1)^{T+S}P=+1
\end{equation}
where T is the isospin, S is the internal spin and P is the dibaryon
parity. These are even singlets and odd triplets at $T=0$,
and odd singlets and even triplets at $T=1$.
The decay of the dibaryons with a symmetric wave function into
two nucleons is suppressed by the Pauli principle.
These dibaryons with the mass \mbox{$M < 2m_{N}+m_{\pi}$} can decay
into two nucleons mainly emitting a photon (real or virtual). The
contribution of such dibaryons to strong interaction processes of hadrons
is very small. However, their contribution to electromagnetic processes
on light nuclei may exceed the cross section for the process under study
out of the dibaryon resonance by several orders of magnitude [18].

In the frame of the MIT bag model Mulders et al [20] calculated the masses
of different dibaryons, in particular, of ones with symmetric wave function.
They predicted dibaryons $D(T=0;J^P =0^{-},1^{-},2^{-};M=2.11\ GeV)$ \ and
$D(1;1^{-};2.2\ GeV)$ corresponding to the states $^{13}P_J$ and $^{31}P_1$.
However, the obtained dibaryon masses exceed the pion production threshold.
Therefore these dibaryons can decay into $\pi NN$ channel. The possibility
of existence of dibaryons with masses $M<2m+m_\pi $ was discussed by
Kondratuk et al [21] in the model of stretched rotating bags with account of
spin-orbital quark interaction. In the frame of the chiral soliton model
Kopeliovich [22] predicted that the masses of $D(1,1^{+})$ and $D(0,2^{+})$
dibaryons exceed the two nucleon mass by 60 and 90 MeV respectively.
These are lower than the pion production threshold.

Unfortunately, all obtained results for the dibaryon masses are very model
dependent. Therefore, only experiment could answer the question about the
existence of dibaryons with symmetric wave function and their masses.

In the present paper we propose the experiment on the search of the dibaryons
with symmetric wave function $D(T=1,J^P=1^-,S=0)$ and $D(1,1^+,1)$
in a processes of a radiative photoproduction of the $\pi^{\pm}$ mesons on
the deuteron
\begin{equation}
\g d \rightarrow \pi^\pm D \rightarrow \pi^{\pm} + \g NN .
\end{equation}
These dibaryons $D(1,1^+,1)$ and $D(1,1^-,0)$
correspond to states $^{3,3}S_1$ and $^{3,1}P_1$ respectively.

\section{The decay widths of the dibaryons $D(1,1^+,1)$ and $D(1,1^-,0)$}\

The decay widths of the dibaryons under consideration into $\g NN$ have been
calculated in [18] assuming that they are determined by a diagram in Fig.1 with
a singlet virtual level $^{31}S_0$ in the intermediate state. Calculations gave
the following expression for the distribution of the probability for the decay
$\dn$ over the energy of an emitted photon $\om$ in the dibaryon rest frame
\begin{equation}
\frac{d \a_{\dn}}{d \om} = \left(\frac{e^2}{4\pi} \right)
\frac{\eta_0 g^2_0 f^2_5}{9\pi^2}\frac{\om^2_0}{M^2}
\frac{\sqrt{2(\om_m- \om)(M- 2\om)}}{(\om_0- \om)^2} \om .
\end{equation}
Here $\om_m=(M^2- 4 m^2_N)/2M$ is the maximum energy of the emitted photon:
$\om_0=(M^2- m^2_0)/2M$ is the photon energy corresponding to the singlet
virtual state with the mass $m_0$; $\eta_0$ is the probability for the full
overlapping of nucleons in the $^{31}S_0$ state. The coupling constant at the
vertex $^{31}S_0 \rightarrow NN$ is equal to
$$
g^2_0=\frac{8 \pi}{m_N |a_s|}
$$
where $a_s$ is the singlet length of the $NN$ scattering.

The decay width $\a_{D(1,1^{+})\rightarrow \g NN}$ is obtained by integrating
the eq. (2) over $\om$:
\begin{equation}
\a_{\gn}=\left(\frac{e^2}{4\pi}\right)\frac{g^2_0}{9\pi^2}
\eta_0 f^2_5\frac{\om^2_0}{M^2}\sqrt{2M} \left[-3\sqrt{\om_m}+
\frac{\om_0+ 2\delta_0}{\sqrt{\delta_0}} \arctan \sqrt{\frac{\om_m}{\delta_0}}
\right]
\end{equation}
where $\delta_0=\om_0-\om_m=2/(M a^2_s)$. The decay width for the dibaryon
$D(1,1^{-},0)$ is determined by means of the relation:
\begin{equation}
\frac{4}{f^2_4} \a_{D(1,1^{-},0)}=\frac{1}{f^2_5} \a_{D(1,1^{+},1)},
\end{equation}
the constant $f^2_{4(5)}$ is associated with a change in the dibaryon quantum
numbers as a result of the photon emission and a transition to the state
$^{31}S_0$.

Table 1 lists the expected decay widths of these dibaryons at various
dibaryon masses $M$ [18].
\begin{table}
\centering
\begin{tabular}{|c|l|l|l|l|l|l|} \hline
$M (GeV)$       & 1.90 & 1.93 & 1.96 & 1.98 & 2.00 & 2.013  \\ \hline
$\a_t(1,1^{+})$ & 0.2  & 2.2  & 8.4  & 16   & 26   & 35  \\
$(eV)$          &      &      &      &      &      &     \\  \hline
$\a_t(1,1^{-})$ & 0.05 & 0.55 & 2.1  & 4    & 6.5  & 8.75 \\
$(eV)$          &      &      &      &      &      &      \\  \hline
\end{tabular}
\caption{ Decay widths of the dibaryons $D(1,1^{+},1)$ and $D(1,1^{-},0)$
at various dibaryon masses $M$. $\a_t \approx \a_{\g NN}$.}
\end{table}

It is evident from eq. (3) that one would expect the appearance of a
narrow peak in the distribution of the probability of the decay
$D(1,1^{\pm})\rightarrow \g NN$ over the photon energy. It is caused
by smallness of the value of $\delta_0$. The calculations showed that
the interval of photon energies from $\om_m$ to $(\om_m-1 MeV)$ contains
about $70\%$ of the contribution to the decay widths of the dibaryons
$D(1,1^{\pm})$. Later at the analysis of the kinematical conditions
we will assume that the energy of the emitted photon does not differ
from $\om_m$ more than by $2 MeV$.

\section{The cross sections of the dibaryon $D(1,1^-,0)$ \newline
photoproduction}\

Let us estimate the cross section of the photoproduction of the dibaryon
$D(1,1^{-},0)$ which, as it is expected, should give the biggest contribution
to this process.
The gauge invariant amplitude of photoproduction of the
dibaryon $D(1,1^-,0)$ may be obtained with help of diagrams in Fig.2.
Such dibaryon could be
produced only if a pion is emitted from the six quark state of the
deuteron. Therefore the vertex of $d\rightarrow \pi D$ is written as
\begin{equation}
F_{d\rightarrow \pi D(1,1^{-},0)}=\frac{g_1}{M} \sqrt{\eta}
\Phi_{\mu \nu} G^{\mu \nu} ,
\end{equation}
where $\Phi_{\mu \nu}=r_{\mu}w_{\nu}-w_{\mu}r_{\nu}$,
$G_{\mu \nu}=p_{\mu}v_{\nu}-v_{\mu}p_{\nu}$, $w$ and $v$ are four-vectors
of the dibaryon and deuteron polarization, $r$ and $p$ are the dibaryon
and deuteron four-momenta, $\eta$ is the probability of occurrence of a
six-quark state in the deuteron.

It worth noting that for the regarded processes the dibaryon with symmetric wave
function may be produced only if the nucleons inside of the deuteron (and
$^{31}S_0$ virtual state) are overlapped quite strongly that a six-quark state
with the deuteron (virtual singlet state) quantum numbers is formed. Therefore
the probabilities of a production and a decay of the dibaryon with
symmetric wave
function have to be proportional to $\eta$ and $\eta_0$ respectively. According
to the estimation of work [23], $\eta=0.03-0.01$. In this paper we assume
$\eta =\eta_0=0.01$. If in the proposed experiment we do not observe the
contribution of the dibaryons larger than 100\% of the background one, then,
in particular, it may indicate that $\eta$ (or $\eta_0$) is smaller than
0.001.

The following matrix elements correspond to the diagrams in Fig.2a,b,c,d:
\beq
T_a&=&-e\sqrt{\eta}\frac{g_1}{M}\frac{(\epsilon (2p_1+k_1))}{(k_1p_1)}
\left\{(vw)(rP)-(vr)(wP)\right\}\nonumber\\
T_b&=&e\sqrt{\eta}\frac{g_1}{M}\frac{(\epsilon (2r-k_1))}{(k_1r)}\left\{
(vw)(p_1Q)-(vQ)(p_1w)\right\}
\nonumber\\
T_c&=&e\sqrt{\eta}\frac{g_1}{M}\frac{(\epsilon (2q_1+k_1))}{(k_1q_1)}
\left\{(vw)(p_1r)-(vr)(p_1w)\right\} \\
T_d&=&T_{d1}+\alpha T_{d2}\nonumber\\
T_{d1}&=&2e\sqrt{\eta}\frac{g_1}{M}\left\{(vw)(\epsilon r)-(vr)(\epsilon w)
\right\}
\nonumber\\
T_{d2}&=&2e\sqrt{\eta}\frac{g_1}{M}\left\{(vw)(\epsilon p_1)-(\epsilon v)
(p_1w)\right\}
\nonumber
\eeq
where $P=(p_1+k_1)$, $Q=(r-k_1)$; $k_1$ and  $q_1$  are  the  four
momenta of
the photon and the pion; $\epsilon$ is a four-vector of the photon
polarization; the coefficient $\alpha$=0,1,2 for production of the $\pP$,
$\pO$, $\pM$ mesons respectively. The expression for the matrix element
$T_d$ is written in the form which ensures the gauge invariance of the
amplitude of the dibaryon photoproduction.

The matrix elements (7) are connected with the amplitudes of the dibaryon
$D(1,1^-,0)$ photoproduction in different channels as:
\beq
T(\g d\rightarrow \pO D)&=&T_a+T_b+T_{d1}+T_{d2} \nonumber\\
T(\dN)&=&\sqrt{2}\left(T_a+T_c+T_{d1}\right) \\
T(\dP)&=&-\sqrt{2}\left(T_a+2T_b-T_c+T_{d1}+T_{d2}\right) \nonumber
\eeq

Let us calculate the cross section of the dibaryon $D(1,1^-,0)$
photoproduction in the reactions with formation of $\pP$ meson. We will use
the calibration $\epsilon_0 =0$. Then the amplitude $T_a=0$ in
lab system $(\epsilon p_1)=0$. As result of calculation for the process
$\dN$ we have (in lab system):
\beq
\lefteqn{\BO =\frac{1}{3}\left(\frac{e^2}{4\pi}\right)\left(\frac{g_1^2}
{4\pi}\right)\eta\frac{q^2}{m_dM^2\nu J}\left\{\frac{1}{2m_d^2}
[(M^2+m_d^2-t)^2-4m_d^2M^2]\right.+} \nonumber \\
&& \left. q^2(1-\cos^2\te)\left[1+2\frac{(M^2+m_d^2-t)^2+2m_d^2M^2}
{(\mu^2-t)^2}-4\frac{M^2+m_d^2-t}{\mu^2-t}\right]\right\}
\eeq
where
$$
s=2m_d\nu+m_d^2, \quad t=\mu^2-2\nu (q_0-q\cos\te), \quad
J=q(m_d+\nu)-q_0\nu \cos\te,
$$
$m_d$ is the deuteron mass, $\nu$ is the incident photon energy, $q_0(q)$
is the $\pi$ meson energy (momentum). The  pion  energy  $q_0$  is
connected with
the pion emission angle $\te$ in the following way:
\be
q_0=\frac{1}{c_1}\left[(m_d+\nu)c_2 \pm \nu\cos\te \sqrt{c_2^2-2\mu^2c_1}
\right]
\ee
where
$$
c_1=2[(m_d+\nu)^2-\nu^2\cos^2\te], \qquad c_2=s+\mu^2-M^2 .
$$

The obtained expression for the cross section of the dibaryon photoproduction
in the reaction $\dP$ is
\beq
\CO&=&\frac{1}{3}\left(\frac{e^2}{4\pi}\right)\left(\frac{g_1^2}
{4\pi}\right)\eta\frac{q^2}{m_d M^2\nu J}
\left\{\frac{1}{2}(M^2+m_d^2-t)^2\left(\frac{1}{m_d^2}+\frac{4}{M^2}\right)
-\right.\nonumber \\
&&\left.(M^2+4m_d^2)+q^2(1-\cos^2\te)\left[1+\frac{A_1}{(M^2-u)^2}+
\frac{A_2}{M^2-u}+\right.\right.\nonumber \\
&&\left.\left.\frac{A_3}{(M^2-u)(\mu^2-t)}+
\frac{A_4}{(\mu^2-t)}+\frac{A_5}{\mu^2-t}^2\right]\right\}
\eeq
where
\beq
A_1&=&4\left[(m_d^2-\mu^2+u)^2+4m_d^2u+ \right.\nonumber \\
&&\left.\frac{M^2+m_d^2-t}{M^2}\left((M^2+u)
(m_d^2-\mu^2+u)-u(M^2+m^2-t)\right)\right], \nonumber \\
A_2&=& 4\left[2(3m_d^2-\mu^2+u)-\frac{1}{M^2}(s-m_d^2)(M^2+m_d^2-t)\right],
\nonumber \\
A_3&=& 8\left[(m_d^2-\mu^2+u)(M^2+m_d^2-t)+m_d^2(M^2+u)\right], \\
A_4&=& 4(M^2+3m_d^2-t), \nonumber \\
A_5&=& 2[(M^2+m_d^2-t)^2+2m_d^2M^2]. \nonumber
\eeq
Here $u=(p_1-q_1)^2=m_d^2+\mu^2-2m_dq_0$.

In the numerical calculations we assumed that $\eta=\eta_0=0.01$.
%Recall that according to modern representation [23], $\eta=0.03 -0.01$.
The coupling constant $g^2_1/4\pi$ is unknown. This is the coupling
constant of the strong interaction. Not to overestimate the value
of the cross section we took it equal to 1.
To estimate the dibaryon contribution at different masses $M$
we will assume the possibility of existance of the dibaryons with masses
$M$=1.9, 1.95 and 2.00 GeV.
Results of the calculations of the differential and total cross sections
of the dibaryon $D(1,1^{-}0)$ production in the reactions $\dN$ and $\dP$
are presented in Fig.3 and 4, respectively. The main contribution to these
processes are given by the diagram in Fig.2c.

Let us consider to carry out a search of the dibaryons at the tagged
photon beam of the microtron MAMI.

\section{Analysis of the dibaryon $D(1,1^-,0)$ production in the reaction
$\gnn$}\

Taking into account a high energy resolution of the tagging system
$(\Delta\nu_1=2MeV)$,  an   optimal   method   of   the   dibaryon
identification
could be the missing mass one. It would allow, at an additional detection
of the final photon, to separate reliably from a background. However
in this case a magnetic spectrometer with high resolution is necessary
to detect charged pions with the kinetic energy 100--600$MeV$.

Another way to recognize the dibaryon is a reconstruction of the
dibaryon mass by detecting the decay products. The information about the
tagged photon energy could be used to suppress background processes.
This method of dibaryon identification is studied in the present paper
in detail.

The Monte-Carlo simulation was used to generate events of the dibaryon
production in the processes (2) and obtain background reactions
\be
\g d\rightarrow \pi^{\pm} + \pO + NN ,
\ee
\be
\g d\rightarrow \pi^{\pm} + \g + NN .
\ee
For simulation of dibaryon photoproduction we used the expression (9)
and (11) for the differential cross sections. Studying the reactions (13)
we used the values of the total cross sections for the reactions
$\g p\rightarrow \pP \pO n$ [25]. The total cross section for the reaction
(14) in an energy region $\nu_1=500-800 MeV$ was took equal to 0.01$\l$,
which is a wittily abnormally high value.

Let us consider in the first a dibaryon production in the reaction
\be
\g d\rightarrow \pP D \rightarrow \pP +\g nn  .
\ee
In this case we must detect the photon and two neutrons.

In order to find the optimal location of the setup, an analysis of
distributions of kinematical variables for investigating reactions
was carried out. These distributions are presented in Fig.5--10
for the dibaryon masses 1900, 1950 and 2000 $MeV$. Fig.5 shows the
distributions over the energy and the emission angle of the neutrons
from the dibaryon decay. The distributions over $\cos\tn$ in this figure
corresponds to an interval of the neutron energy from 10 to 100$MeV$.
The average angle of the neutron detector location is choose from these
distributions to be equal to $45^{\circ}$. The choice of this angle
in the side of bigger dibaryon mass is caused by the bigger cross section
of the production of the dibaryon with smaller mass.

Fig.6 demonstrates distributions over the energy of the photon from the
dibaryon decay and over $\cos\theta_{\g n}$, where $\theta_{\g n}$ is
an angle between the neutron and the final photon. These distributions were
obtained also for the neutron energy in the interval 10--100$MeV$.
It is evident from these distributions that the optimal angle of the
$\g$ detector location depends on the dibaryon mass. For $M=1900 MeV$
this angle coincides with the angle of the location of the neutron detector.
For $M=2000 MeV$ the optimal location of the $\g$ detector is opposite
to the neutron detector. Taking into account increase of the cross section
at decrease of the dibaryon mass we chose the latter position of the $\g$
detector $(\tg=135^{\circ})$.

As the neutron detector we shall consider the time-of-flight scintillator
detector consisted of 5 planes with sizes $300\times 300\times 5\; cm^3$.
Each plane contains from 15 scintillator counters with sizes $300\times
20\times  5\;  cm^3$.  This  detector  has  a  neutron   detection
efficiency
20--30\% in the energy interval 10--100$MeV$.

As a $\g$ detector we shall  assume  the  block  from  64  $BaF_2$
crystal
with sizes $40\times 50\times 24 \ cm^3$. The parameters of the setup and
the location of the $\g$ and neutron detectors are listed in Table 2.
\begin{table}
\centering
\begin{tabular}{|c|c|c|c|c|} \hline
detectors & $\theta$ & $\varphi$ & $R(cm)$ & sizes $(cm^2)$ \\ \hline
$\g$      &          &           &         &$40\times 50 \times 25$  \\
$(BaF_2)$ &$235^{\circ}$&$270^{\circ}$&100&  (segmented -- 64) \\ \hline
$n$       &        &         &         &$300\times 300\times 5$  \\
(TOF)     &$45^{\circ}$&$90^{\circ}$&400& (5 planes) \\ \hline
\end{tabular}
\caption{ Parameters and location of the $\g$ and neutron detectors.}
\end{table}

It was simulated quantity of events for the processes (2), (13) and (14)
corresponding to 1 hour of the exposition time of the microtron MAMI
for the energy interval of the tagged photons 500--800$MeV$, the incident
photon intensity $1.3\cdot 10^6 sec^{-1}$ at $\nu_1=500 MeV$ in the interval
$\Delta\nu_1=2 MeV$ and a thickness of the deuteron target 10$cm$.
Obtained a geometrical efficiency at the different values of the dibaryon mass
are presented in Table 3.
\begin{table}
\centering
\begin{tabular}{|c|c|} \hline
$M\ (MeV)$  & $\varepsilon \ (\%)$   \\ \hline
 1900       &      0.05           \\ \hline
 1950       &      0.09           \\ \hline
 2000       &      0.16           \\ \hline
\end{tabular}
\caption{ Geometrical efficiency of the setup at different values of $M$.}
\end{table}

Increase of the geometrical efficiency of the setup for the dibaryon with
bigger mass is well founded as increase of the dibaryon mass results in
growth of dibaryon identification errors and the contribution of the
background processes.

Let us consider kinematical parameters of two neutrons from the dibaryon decay.
The kinetic energy of these neutrons in cms are close to zero (0.07--1.9
$MeV$ for $M=1900-2000\ MeV$). In lab.syst. the emission angles and the
energies of neutrons are determined mainly by motion of the cms. Both neutrons
in lab.syst. have the close emission angles. Fig.7b,e,h show the distributions
over $\cos\theta_{nn}$, where $\theta_{nn}$ is the angle between the emission
directions of these neutrons. It is evident that $\theta_{nn}\le 15^{\circ}$.

The average neutron emission angle in lab.syst. coincides practically with
the emission angle of the dibaryon. In this case the photons, emitted to
the opposite direction, have minimum values of the energy (the energy spectra
for such photons are shown in Fig.7a,d,g) and so minimum values of the
absolute error of this energy measurement.

In Fig.7c,f,i the distributions over $\delta T_{nn}=2|T_{n1}-T_{n2}|/(T_{n1}
+T_{n2})$ are presented. These distributions show that the energies of two
neutrons differ by smaller than 50\%. To detect such pairs of neutrons,
high enough discretness of the neutron detector is necessary.

The carried out simulation resulted in the expected yields of the reactions
(13)--(15) for 1 hour of the exposition time of the microtron MAMI for
described setup (Table 4).
\small
\begin{table}
\centering
\begin{tabular}{|c|r|r|r|r|} \hline
           & M=1.9GeV    & M=1.95GeV    & M=2.00GeV    & background \\ \hline
simulated  &             &              &              &            \\
 events    & 421400      &  348200      &  277800      &17897900    \\ \hline
$T_n$=10-100MeV& 229400  &  200700      &  181800      & 5010700    \\ \hline
$nn$       &   25700     &  27100       &   23100      & 108900     \\ \hline
$\g nn$    &    202      &   307        &     444      &   736      \\ \hline
M=1870-2040MeV& 202      &   307        &     444      &   636      \\ \hline
\end{tabular}
\caption{ The expected yields of the dibaryon production in the process
$\g d\rightarrow \pP +D\rightarrow \pP +\g nn$ and the background reactions
(13) and (14) for 1 hour of exposition time of the microtron MAMI.}
\end{table}
\normalsize
The first line of Table 4 shows the total numbers of the events. The
following lines demonstrate numbers of events left after using limitation
of the value of the neutron energy, two neutron coincidence condition, triple
coincidences $\g nn$ and limited range of the dibaryon masses.

The yields of the dibaryon production are expected to be 200--400(1/hour)
for the dibaryon masses in the interval $M=1900-2000 MeV$. If take into
account that the efficiency of detection of two neutrons by the neutron
detector is about 0.05\%, then to obtain such yields really it is necessary
to have 20 hours of the exposition time. The expected spectra of the
dibaryon mass for $M$=1900, 1950, 2000 $MeV$ are shown in the Fig.8a (without
background) and Fig.8b (with the background from the reactions (13) and (14)).

We did not take into consideration the background from random coincidences and
from the reaction
\be
\g + d\rightarrow \pO np
\ee
In principle, the detection of the neutron from this reaction in two blocks of
TOF detector could imitate the detection of two neutrons. Besides, it is
possible to detect a photon from the $\pO$ meson decay. The contribution of
this background should be considered. However, preliminary estimations
indicate that this contribution does not exceed one from the process (13).

Taking into account all mentioned above, it needs about 100 hours of
exposition time to obtain unambiguous result.

%\newpage
\section{Analysis of the dibaryon $D(1,1^-,0)$ production in the reaction
$\gp$}\

The analogous simulations were carried out for the process
\be
\g +d\rightarrow \pM +D\rightarrow \pM +\g pp
\ee
In this case we must detect the photon and two protons. To detect the photon
it is assumed to use the $\g$ detector such as above. To detect the protons
it could be used a time-of-flight scintillator spectrometer consisted of
3 planes  with  sizes  $300\times  300\times  5cm^3$.  Each  plane
contains 15
scintillator blocks with sizes $300\times 20 \times 5cm^3$. To limit a range
of charged particles, this spectrometer should be supplemented by a plane
of the anticoincident counters behind of the TOF detector.

In order to identify the protons the relation of range--energy and the
time of flight of the detected particle will be used.

Results of the kinematical investigation for the reaction (17) are presented in
Fig.9--11. Fig.9a,c,e show that the proton energy does not exceed 150$MeV$.
The range of a proton with such an energy corresponds to the thickness of
the considering spectrometer $(15cm)$. This spectrometer allows to detect
protons with the energy higher than $\sim 50MeV$. Protons with lower energy
either are absorbed or are strongly scattered in the target and air.

The angular distributions of the protons with energy 50--150$MeV$ are shown
in Fig.9b,d,f. These distributions give the average angle of the proton
detector location equal to $35^{\circ}$.

Fig.10 demonstrates the energy and angular distributions of the photons from
the reaction (17) when the protons have energy in the region of 50--150$MeV$.
Using the same arguments as in the previous case we chose the angle of the
$\g$ detector location $145^{\circ}$.

Fig.11 shows the distributions over the photon energy, the angle between
two protons $\theta_{pp}$ and $\delta T_{pp}=|T_{p1}-T_{p2}|/(T_{p1}+T_{p2})$
for different dibaryon masses at the chosen location of the detectors.
Fig.11a,d,g indicate (see also Fig.7a,d,g) that the narrow peak in the photon
spectrum can be an additional signal of the dibaryon production.

The expected yields of the dibaryon production in the reaction (17) are
equal to $\sim 150$/hour for each dibaryon mass. The expected spectra of
mass for three dibaryon masses are presented in Fig.12a (without background)
and Fig.12b (with the background).

The main difficulty of investigation of the reaction (17) is identification
and determination of energy of two protons flying at close angles.
It can decrease the efficiency of the setup and make worse the resolution over
mass. Moreover it is necessary to consider the random background contribution.
All these could lead to increase of the exposition time.

We propose to investigate the process (17) simultaneously with the process
(15), where it is necessary to have $\sim 100$ hours of the exposition time.
This time would be quite enough for research of the process (17) too. The
proposed location of the setup in this case is  shown  in  Fig.13.
Investigation
both of these reactions allows answer a question about existance of the
dibaryons with symmetric wave function more unambiguously and to decrease
the systematic experimental errors.

\section{Conclusion}\

We studied
the possibility of observation of the supernarrow dibaryons with the symmetric
wave function in the reactions of charged pion photoproduction.
The method of the dibaryon identification by detecting the dibaryon decay
products was analyzed in detail. The cross sections of the dibaryon
photoproduction were calculated. To find the optimal location of the detectors
the kinematical analysis was used. The contributions of the main background
processes were estimated and the expected yields of the dibaryon production
were calculated. The calculations showed that 100 hours of the exposition time
of the microtron MAMI (Mainz) would be quite enough to determine whether or
not the supernarrow dibaryons with the symmetric wave function exist.
It was indicated that the additional signal of the dibaryon production can
be the narrow peak in the photon spectrum.

If in this experiment we does not observe the contribution of the dibaryons
larger than 100\% of the background one, then, in particular, it may indicate
that the probability of six-quark state inside of the deuteron (or $^{31}S_0$
state) is smaller than 0.001.

The suggested search of the dibaryon production in the both (15) and (17)
reactions simultaneously would allow to make more correct conclusion about
such dibaryons existance and decrease the systematical experimental errors.

\section*{Acknowledgments}\

We thank J.Ahrens, D,Drechel, J.Friedrich, N.Nikolaev and Th.Walcher
for stimulating discussion. This work was supported in part by the Russian
Fund for Fundamental Research (Project number: 96-02-16530-A).

\newpage
{\bf References}

\vspace{1cm}

\begin{enumerate}
\item R.L. Jaffe, Phys.Rev.Lett. {\bf 38} (1977) 195; P.J.G. Mulders,
A.T. Aerts and J.J. de Swart, Phys.Rev.Lett. {\bf 40} (1978) 1543,
Phys.Rev. {\bf D21} (1980) 2653; D.B. Lichtenberg {\em et al}.,
Phys.Rev. {\bf D18} (1978) 2569; V. Matveev and P. Sorba,
Lett.Nuovo Cim. {\bf 20} (1977) 425.
\item A.A. Bairamov {\em et al}., Sov.J.Nucl.Phys. {\bf 39} (1984) 26.
\item S.A. Azimov {\em et al}., Sov.J.Nucl.Phys. {\bf 42} (1985) 579.
\item O.B. Abdinov {\em et al}., Sov.J.Nucl,Phys. {\bf 44} (1986) 978.
\item B. Bock {\em et al}., Nucl.Phys. {\bf A459} (1986) 573.
\item O.B. Abdinov {\em et al}., JINR preprint PI-88-102.
\item B. Tatischeff {\em et al}., Z.Phys. {\bf A328} (1987) 147.
\item V.P. Andreev {\em et al}., Z.Phys. {\bf A327} (1987) 363.
\item B. Tatischeff {\em et al}., Relativistic Nuclear Physics and
Quantum Chromodynamics (Proc. of the IX Intern. Seminar on High Energy
Physics Problems, Dubna, Russia, 1988), Dubna, JINR, (1988), p.317.
\item Yu.A Troyan, {\em et al}., Sov.J.Nucl.Phys. {\bf 54,} (1991) 1301.
\item V.V. Glagolev {\em et al}., JINR preprint EI-89-246.
\item E.N. Komarov, Relativistic Nuclear Physics and Quantum
Chromodynamics (Proc. of the XI Intern. Seminar on High Energy Physics
Problems, Dubna, Russia, 1994), Dubna, JINR, (1994), p.321.
\item K.K. Seth, Int. Conf. on medium and high energy physics,
 Taipei, Taiwan, 1988.
\item L.V. Fil'kov, Sov.Physics---Lebedev Inst. Reports No. 11 (1986) 49.
\item L.V. Fil'kov, Sov.J.Nucl.Phys. {\bf 47} (1988) 437.
\item D.M. Akhmedov {\em et al}., Proceed. of the 8th Seminar
``Electromagnetic interactions of nuclei at low and medium energies'',
Moscow, (1991), p.228.
\item D.M. Akhmedov {\em et al., ibid\/}, p.252.
\item D.M. Akhmedov and L.V. Fil'kov, Nucl.Phys. {\bf A544} (1992) 692.
\item S.B. Gerasimov and A.S. Khrykin, JINR Rapid Communication
No. 61571-92.
\item P.J.G. Mulders {\em et al}., Phys. Rev. D {\bf 19} (1979) 2635.
\item L.A. Kondratuk {\em et al}., Sov.J.Nucl.Phys. {\bf 45,} (1987) 1252.
\item V.B. Kopeliovich, Sov.J.Nucl.Phys. {\bf 58,} (1995) 1317.
\item L.A. Kondratyuk {\em et al}., Yad.Fiz. {\bf 43} (1986) 1396.
\item M. Anghinolfi {\em et al}., Research Summaries of the Gordon
Conference 1994 on Photo-Nuclear Physics, p.149.
\item M. Maccormick {\em et al}. (DAPHNE collaboration), Phys.Rev. C,
to be published.

\end{enumerate}

\normalsize
\pagestyle{plain}
\setcounter{figure}{0}
%---------------------------Fig1
\newpage
\begin{figure}[htp]
\epsfxsize=10.cm\epsfysize=5.cm
\hspace*{2cm}
\epsffile{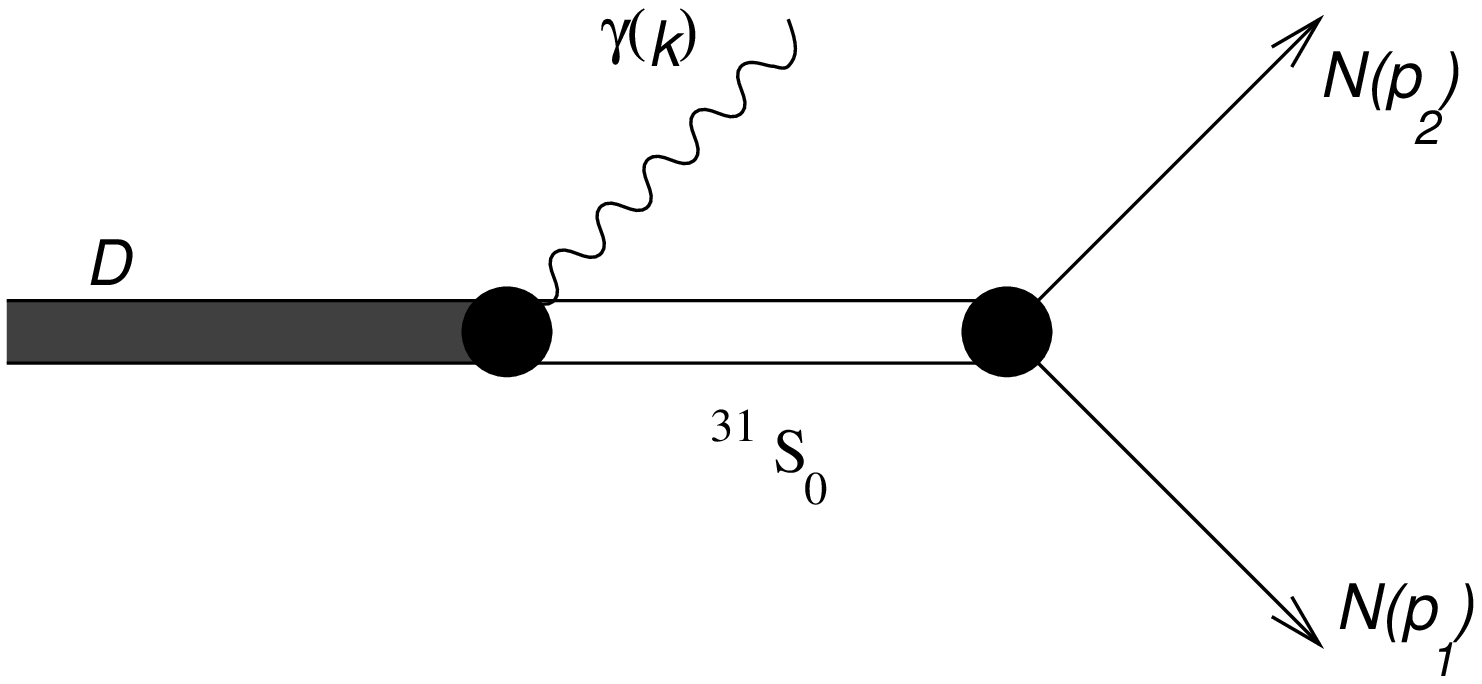}
\caption{The diagram of the dibaryon decay into $\g NN$.}
%---------------------------Fig2
\vspace*{4cm}
\epsfxsize=12.5cm\epsfysize=8.cm
\epsffile{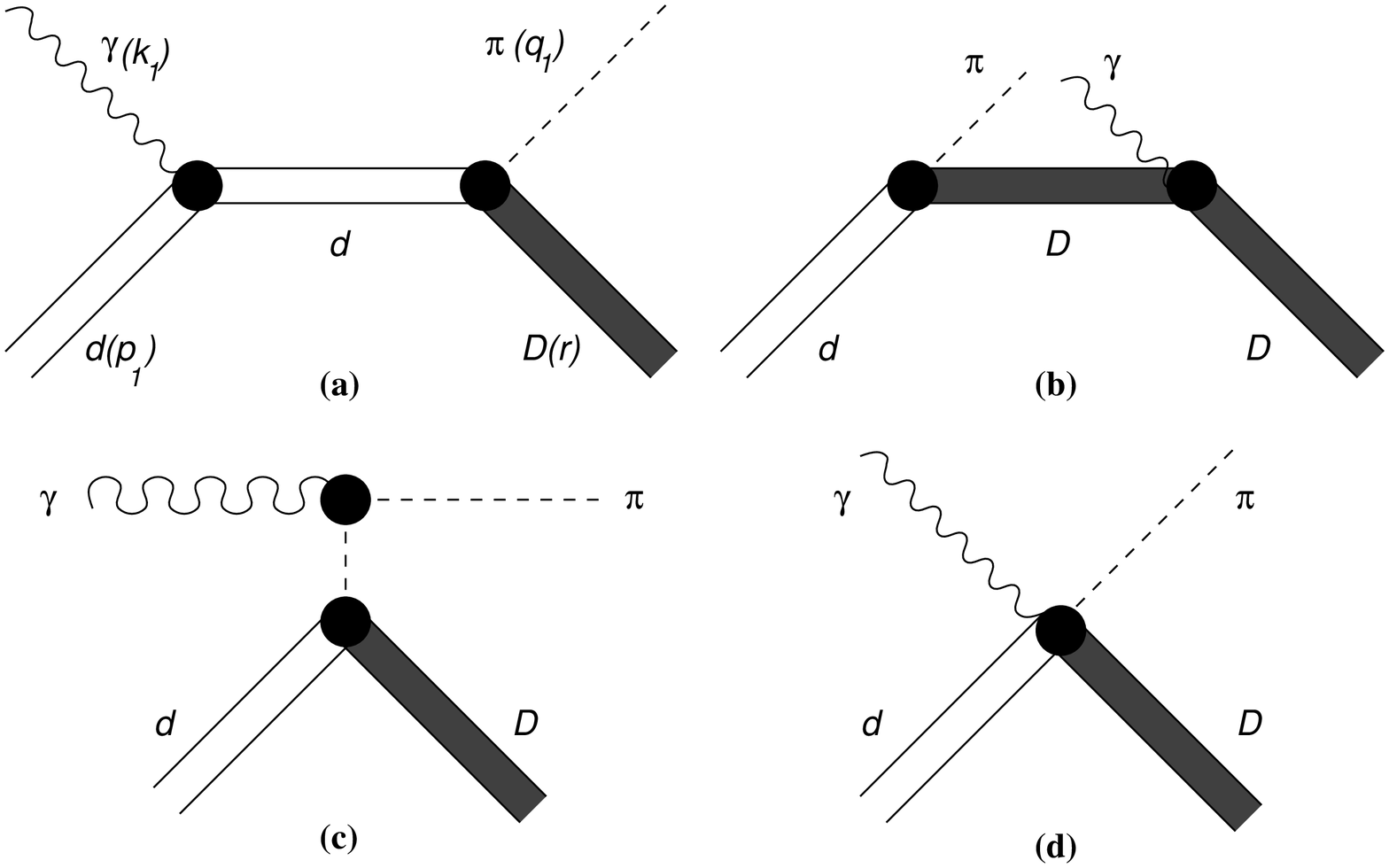}
\caption{The diagrams of the dibaryon production in the process
$\g d\rightarrow \pi D$.}
\end{figure}
%---------------------------Fig3
\newpage
\begin{figure}[htp]
\hspace*{-0.7cm}
\epsfxsize=18.5cm\epsfysize=20.5cm
\epsffile{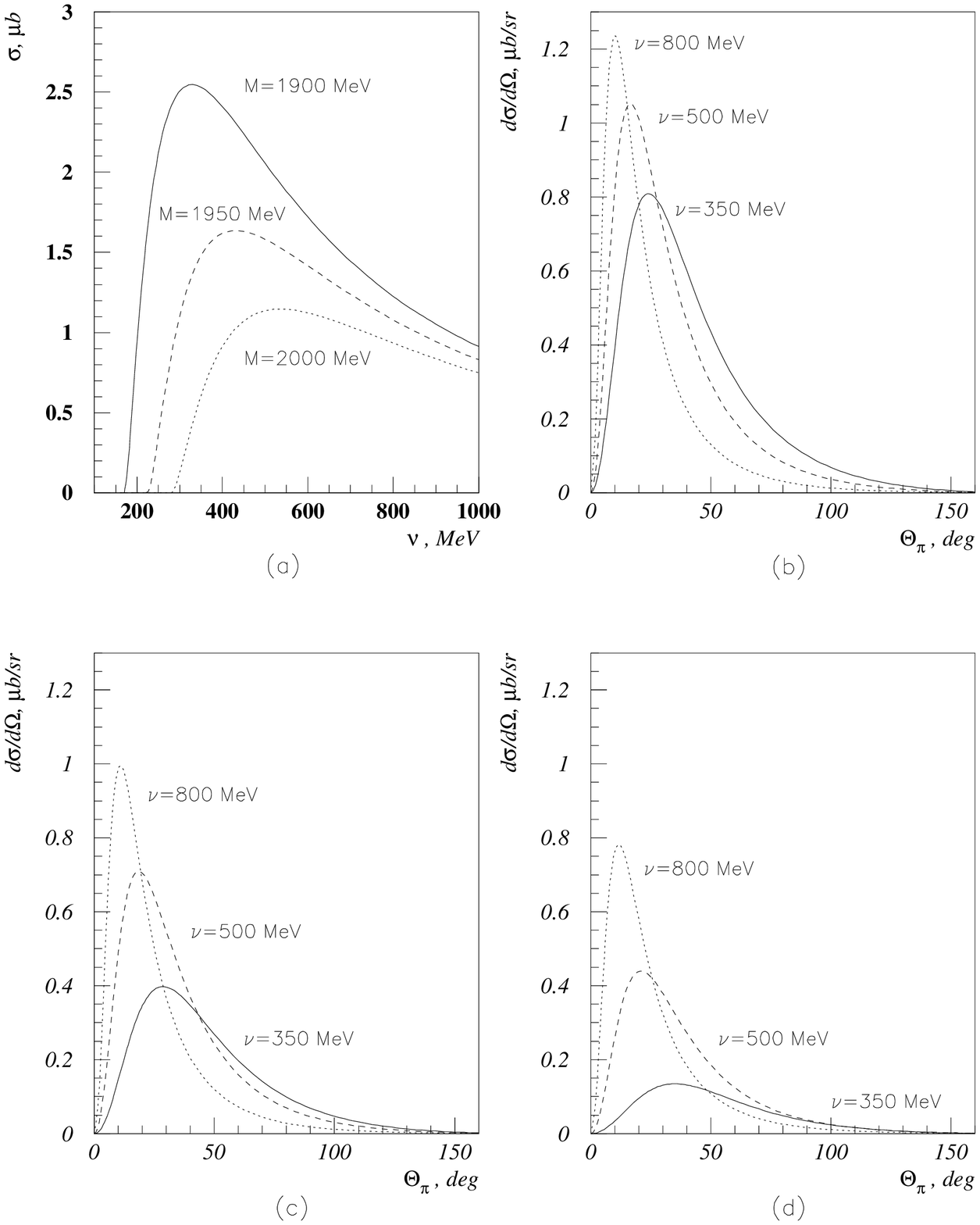}
\vspace*{-1.5cm}
\caption{The cross sections of the dibaryon $D(1,1^-,0)$ production
in the reaction $\dN$ for the different dibaryon masses;
a)--the total cross sections; b,c,d)--the differential cross sections
for M=1900, 1950 and 2000 MeV consecutively.}
\end{figure}
%---------------------------Fig4
\newpage
\begin{figure}[htp]
\hspace*{-0.7cm}
\epsfxsize=18.5cm\epsfysize=20.5cm
\epsffile{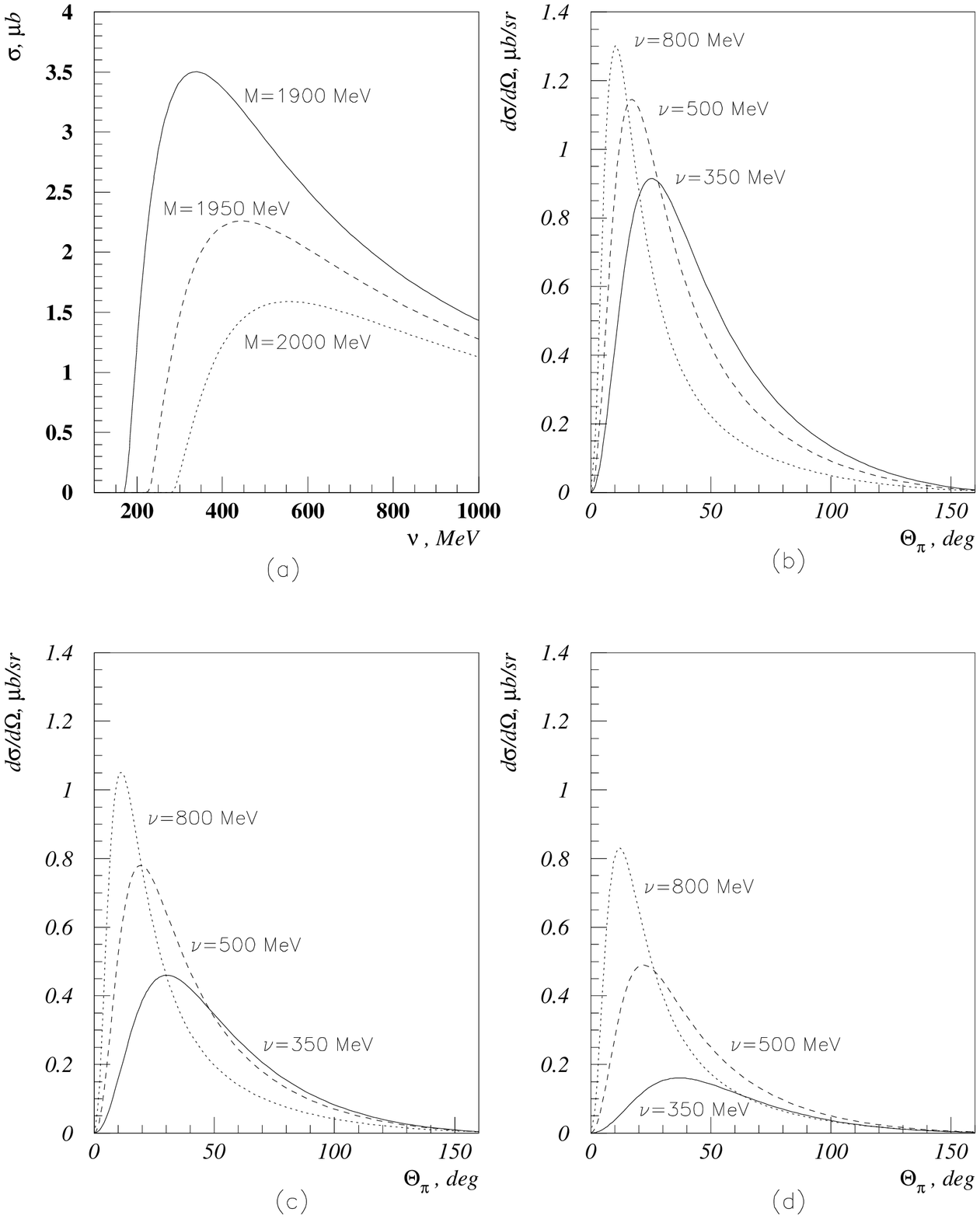}
\vspace*{-1.5cm}
\caption{The cross sections
of the dibaryon $D(1,1^-,0)$ production in the reaction $\dP$; a)--the total
cross sections; b,c,d)--the differential cross sections
for M=1900, 1950 and 2000 MeV consecutively.}
\end{figure}
%---------------------------Fig5
\newpage
\begin{figure}[htp]
\hspace*{-0.7cm}
\epsfxsize=18.5cm\epsfysize=20.5cm
\epsffile{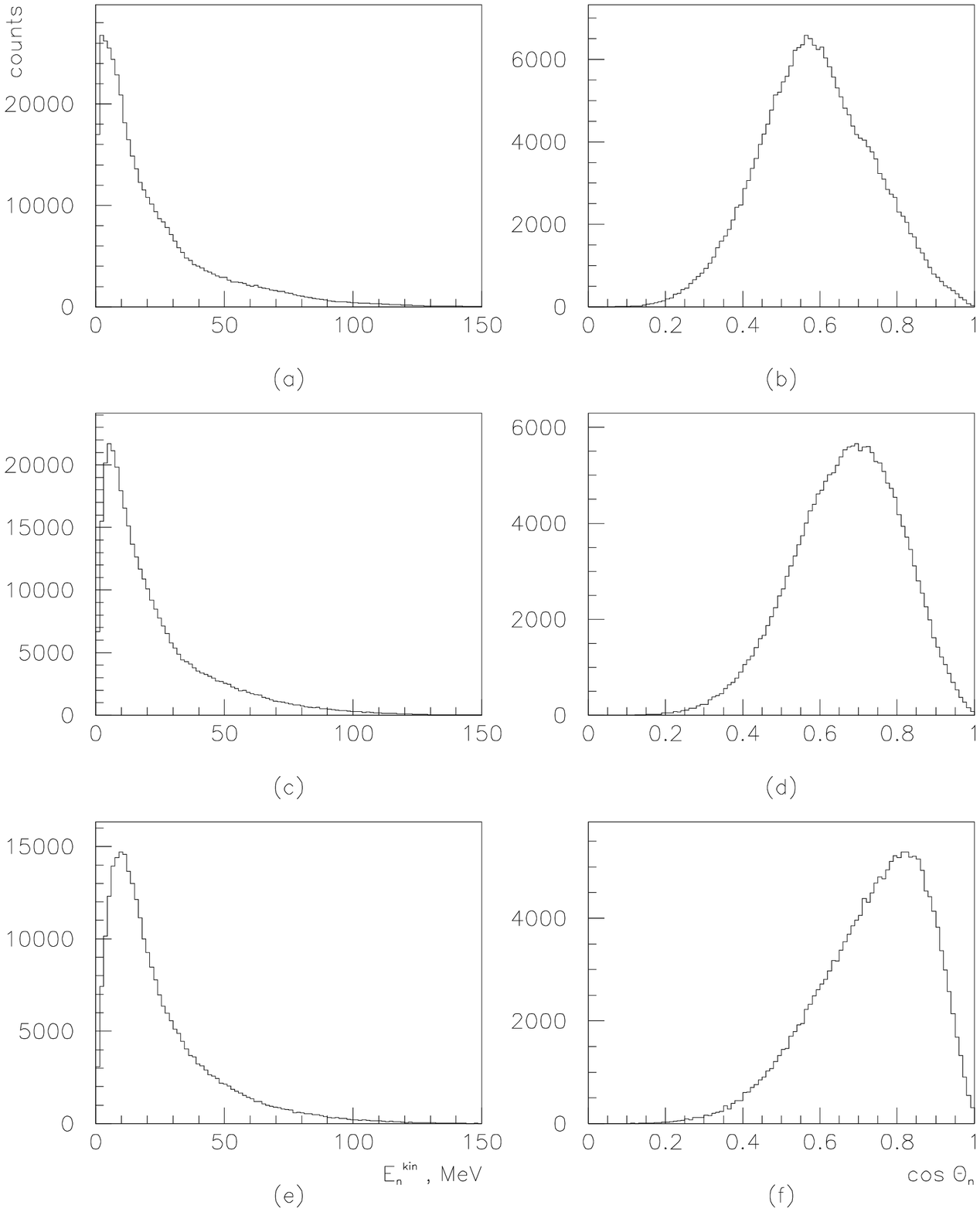}
\vspace*{-1.5cm}
\caption{The energy (a,c,e) and angular (b,d,f) distributions of
the neutrons from the decays of the dibaryons with different masses:
a,b)- M=1900 MeV, c,d)- M=1950 MeV and e,f)- M=2000 MeV.}
\end{figure}
%---------------------------Fig6
\newpage
\begin{figure}[htp]
\hspace*{-0.7cm}
\epsfxsize=18.5cm\epsfysize=20.5cm
\epsffile{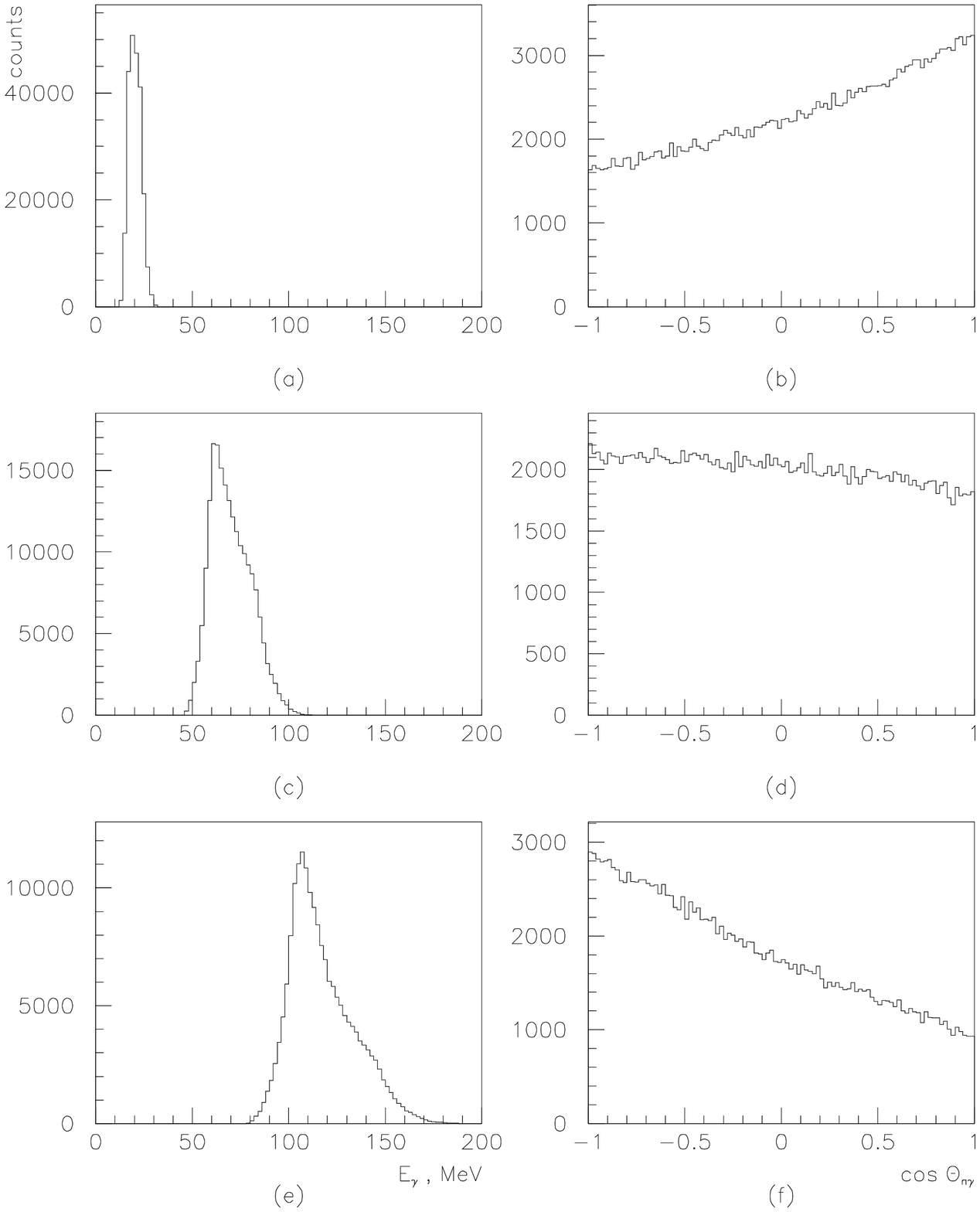}
\vspace*{-1.5cm}
\caption{The energy (a,c,e) and angular (b,d,f) distributions of the
photons from the decays of the dibaryons produced
in the reaction $\dN$ with different masses:
a,b)- M=1900 MeV, c,d)- M=1950 MeV and e,f)- M=2000 MeV.}
\end{figure}
%---------------------------Fig7
\newpage
\begin{figure}[htp]
\hspace*{-0.7cm}
\epsfxsize=18.5cm\epsfysize=20.5cm
\epsffile{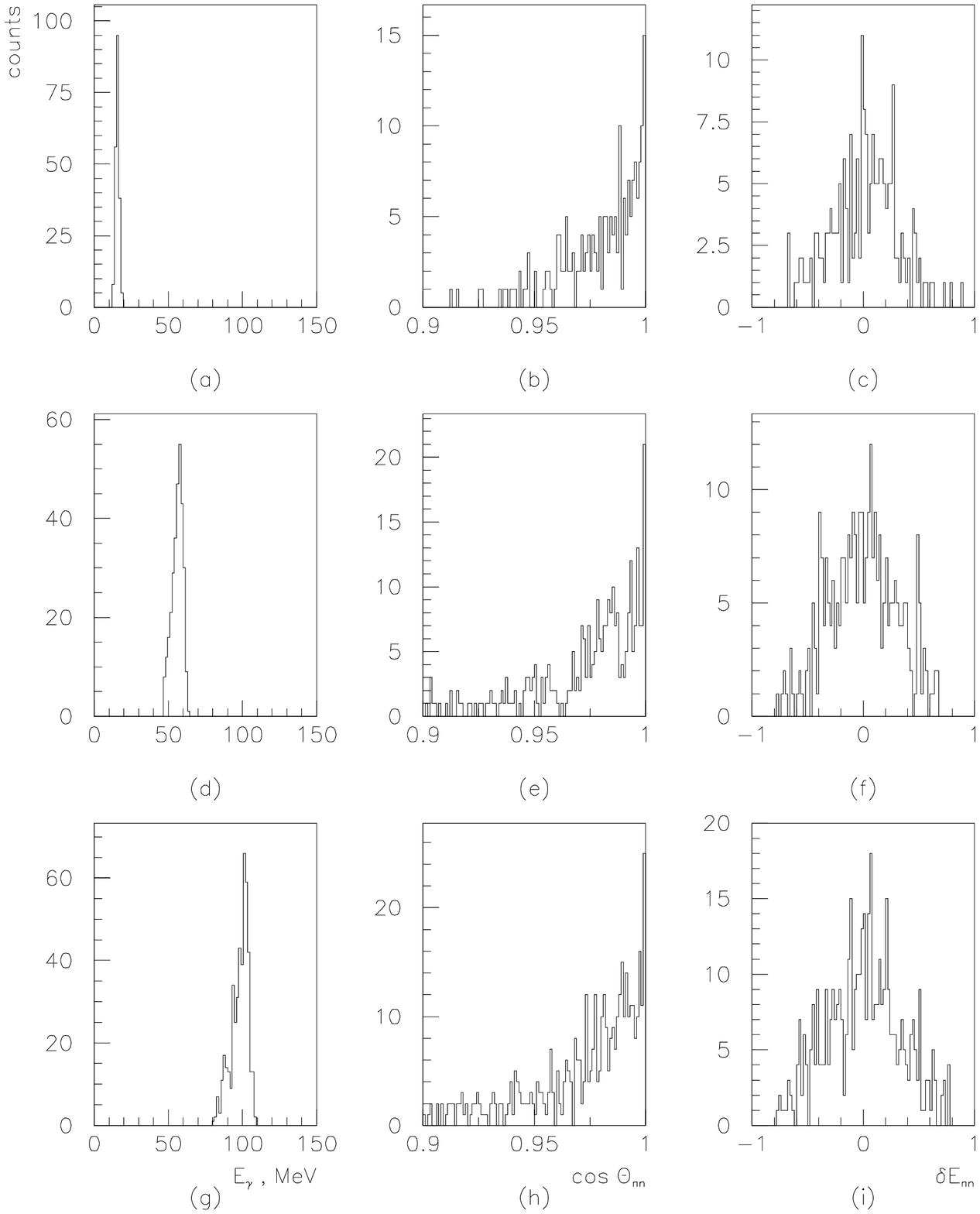}
\vspace*{-1.5cm}
\caption{a,d,g)-- energy spectra of the photons detected by the setup
at the triple coincidences condition $(\g nn)$; b,e,h)-- the distributions over
$\cos\theta_{nn}$; c,f,i)-- the distributions over $\delta T_{nn}$
for different masses:
a,b,c)- M=1900 MeV, d,e,f)- M=1950 MeV and g,h,i)- M=2000 MeV.}
\end{figure}
%---------------------------Fig8
\newpage
\begin{figure}[htp]
\hspace*{-0.9cm}
\epsfxsize=18.5cm\epsfysize=20.5cm
\epsffile{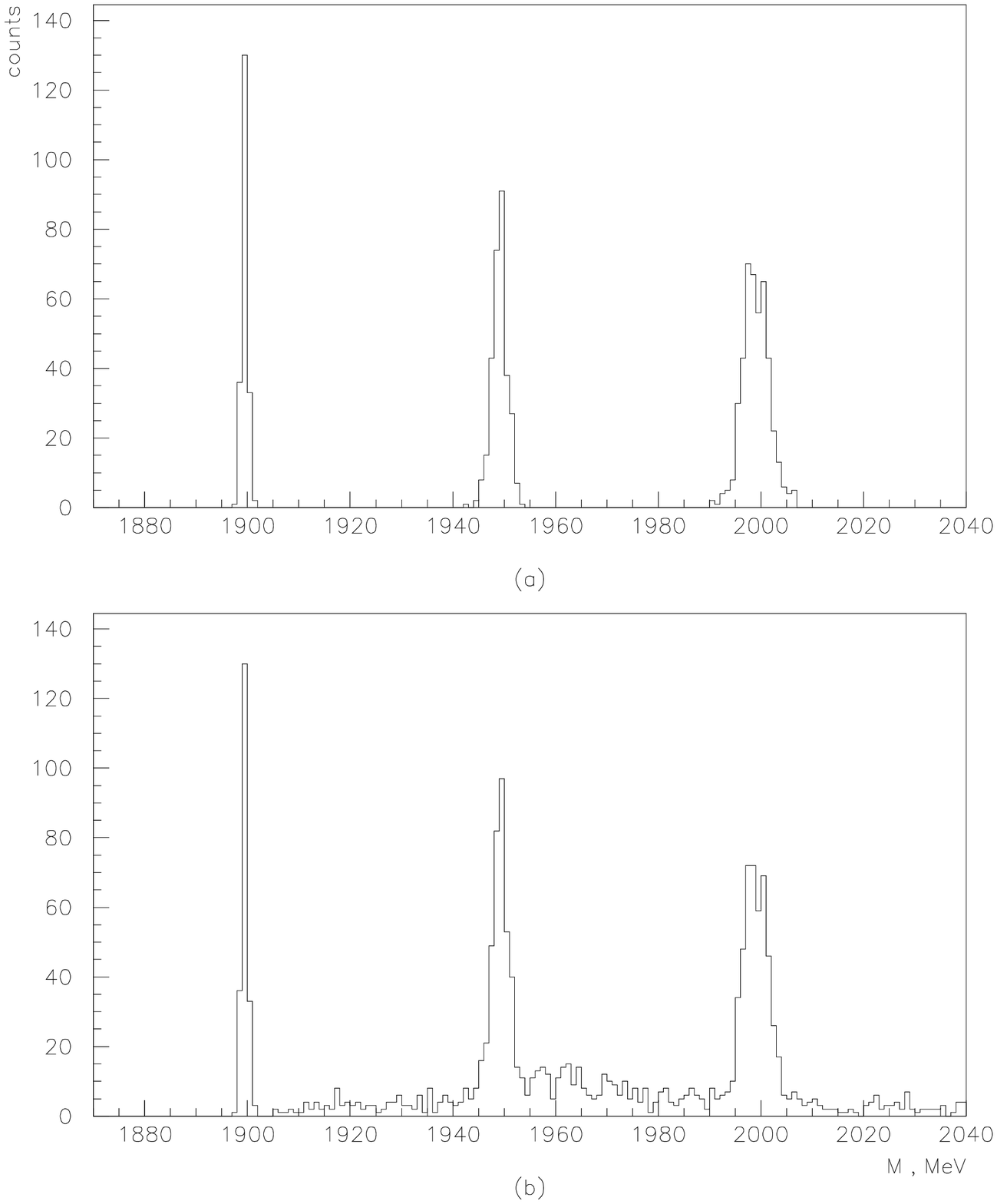}
\vspace*{-1.5cm}
\caption{The expected spectra of mass of the dibaryons produced in
the reaction $\dN$ with the masses $M$=1900, 1950 and 2000$\ MeV$; a)-- without
background contribution; b)--with the contribution of the background from
the processes (13) and (14).}
\end{figure}
%---------------------------Fig9
\newpage
\begin{figure}[htp]
\hspace*{-0.7cm}
\epsfxsize=18.5cm\epsfysize=20.5cm
\epsffile{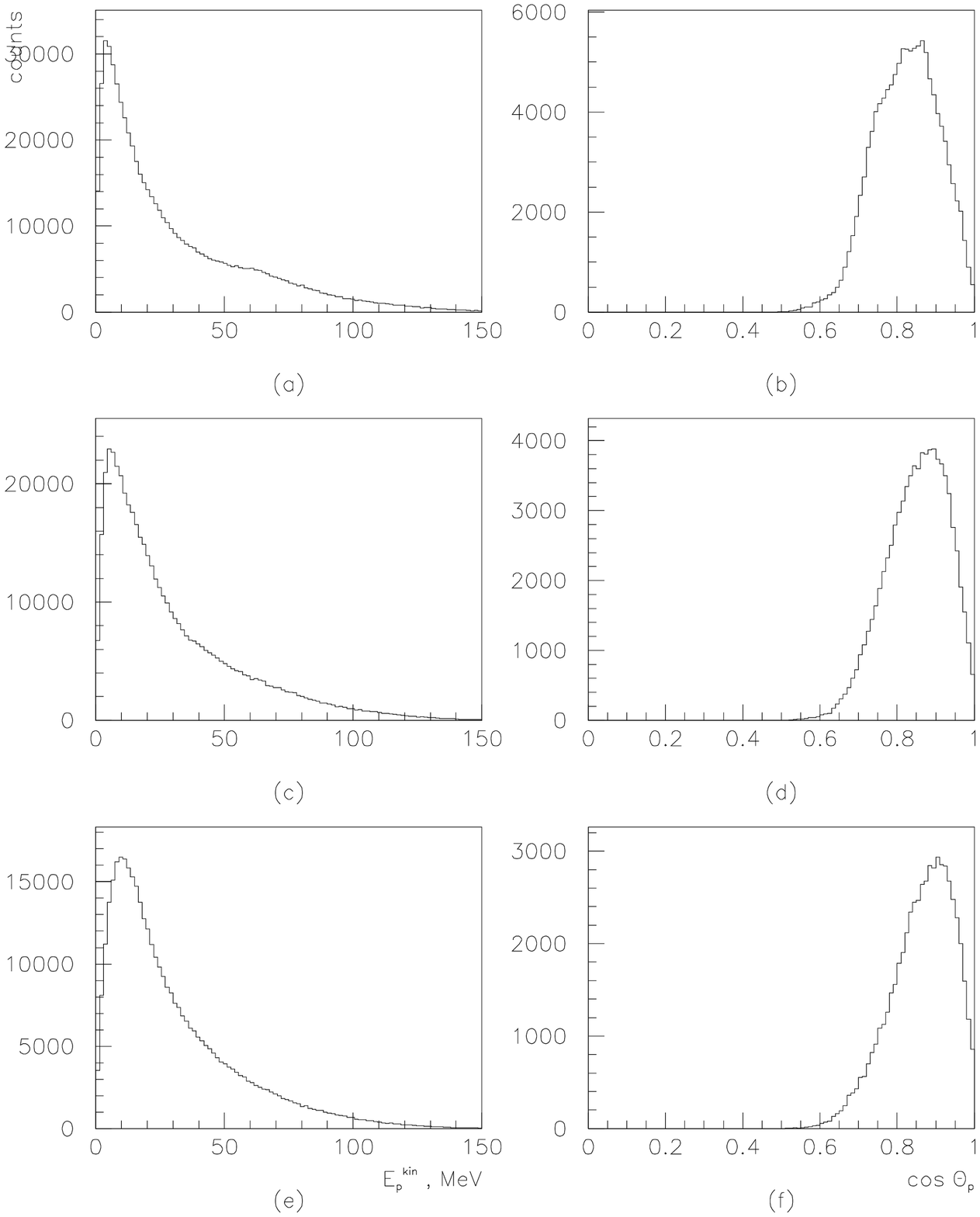}
\vspace*{-1.5cm}
\caption{The energy (a,c,e) and angular (b,d,f) distributions of
the protons from the decays of the dibaryons with different masses:
a,b)- M=1900 MeV, c,d)- M=1950 MeV and e,f)- M=2000 MeV.}
\end{figure}
%---------------------------Fig10
\newpage
\begin{figure}[htp]
\hspace*{-0.7cm}
\epsfxsize=18.5cm\epsfysize=20.5cm
\epsffile{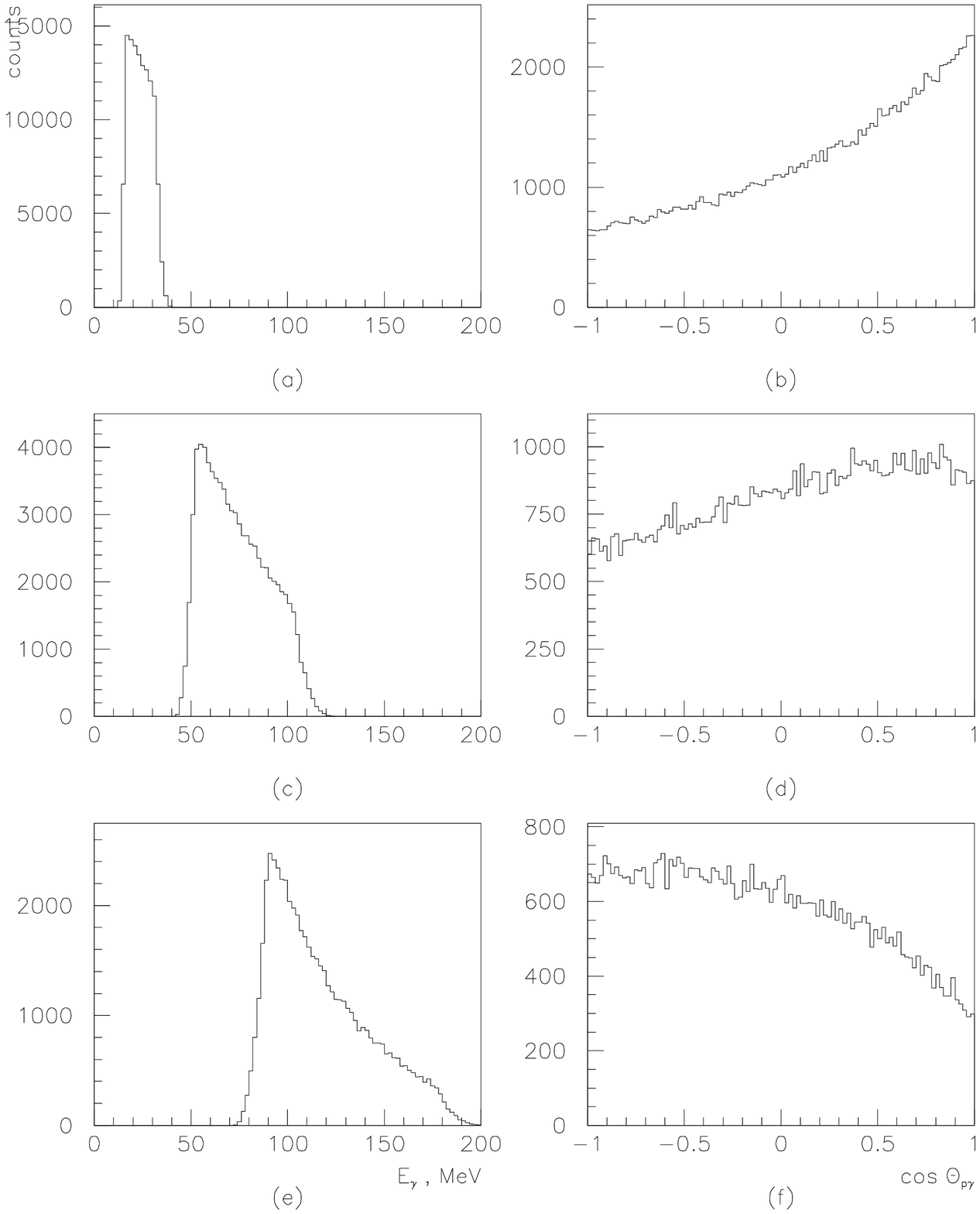}
\vspace*{-1.5cm}
\caption{The energy (a,c,e) and angular (b,d,f) distributions of the
photons from the decays of the dibaryons produced in the reaction $\dP$
with different masses:
a,b)- M=1900 MeV, c,d)- M=1950 MeV and e,f)- M=2000 MeV.}
\end{figure}
%---------------------------Fig11
\newpage
\begin{figure}[htp]
\hspace*{-0.9cm}
\epsfxsize=18.5cm\epsfysize=20.5cm
\epsffile{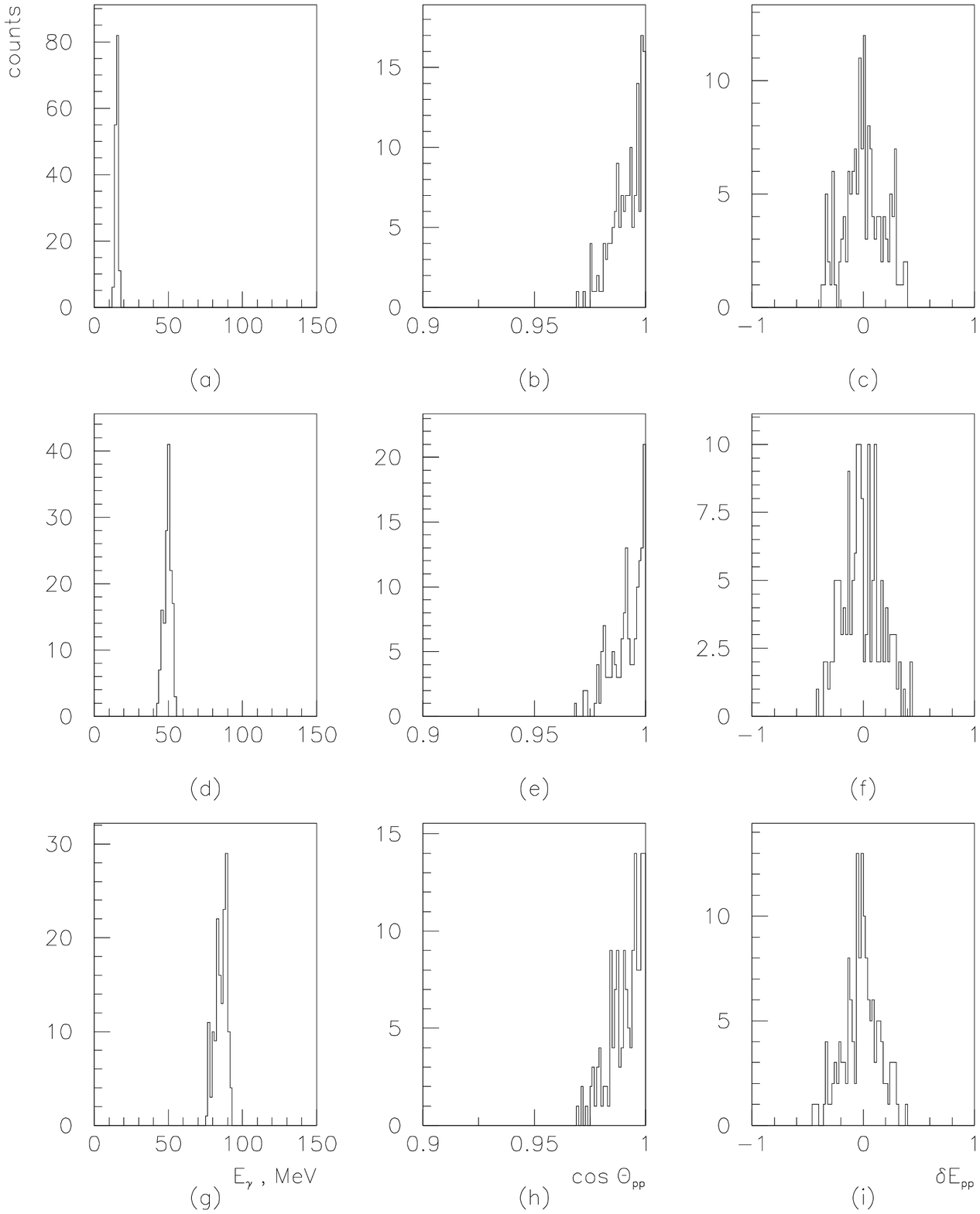}
\vspace*{-1.5cm}
\caption{a,d,g)-- energy spectra of the photons detected by the setup
at the triple coincidences conditions $(\g pp)$; b,e,h)--the distributions over
$\cos\theta_{pp}$; c,f,i)-- the distributions over $\delta T_{pp}$
for different masses:
a,b,c)- M=1900 MeV, d,e,f)- M=1950 MeV and g,h,i)- M=2000 MeV.}
\end{figure}
%---------------------------Fig12
\newpage
\begin{figure}[htp]
\hspace*{-0.9cm}
\epsfxsize=18.5cm\epsfysize=20.5cm
\epsffile{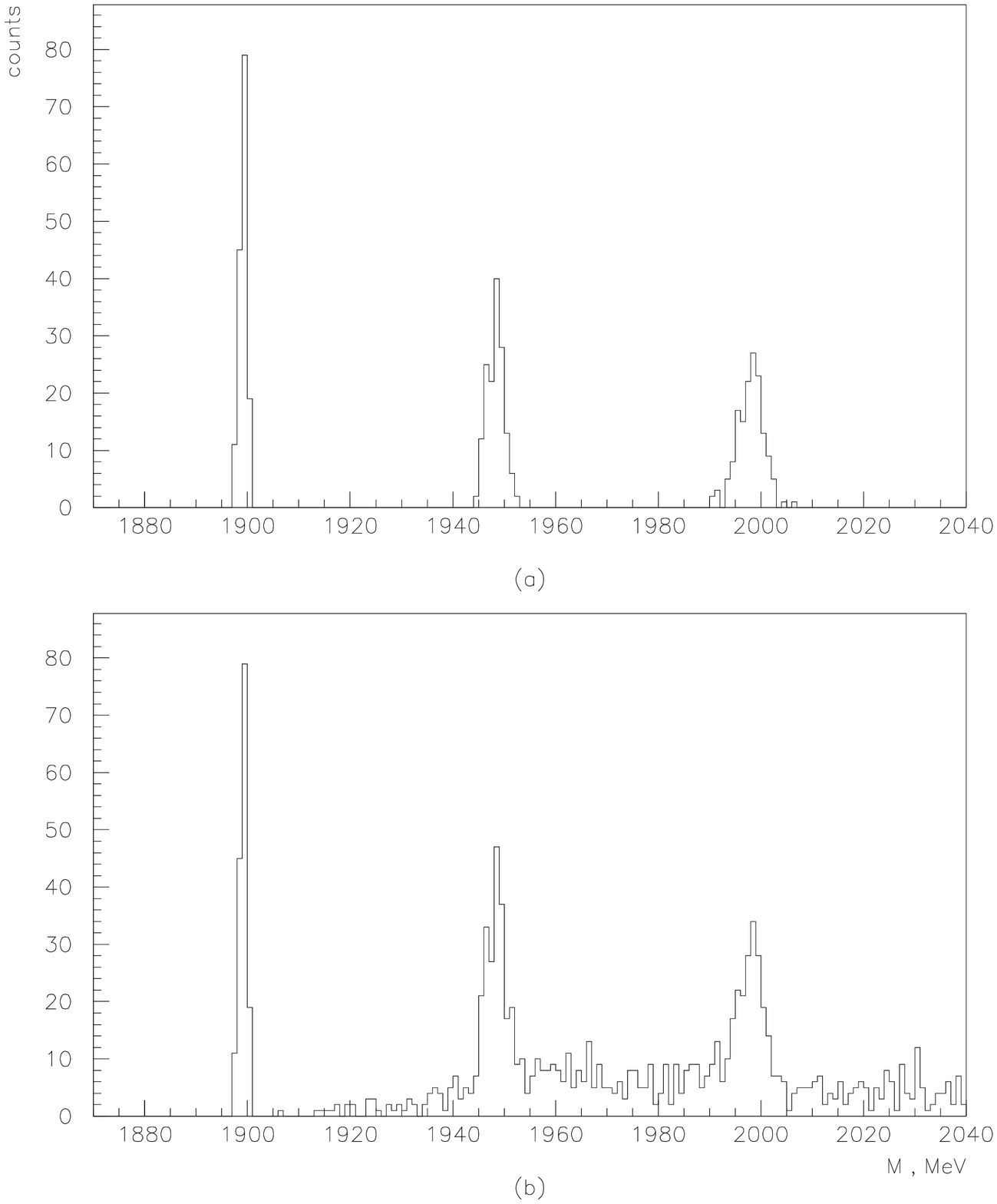}
\vspace*{-1.5cm}
\caption{The expected spectra of mass of the dibaryons produced in the
reaction $\dP$ with the masses $M$=1900, 1950 and 2000$\ MeV$; a)-- without
background contribution; b)-- with the contribution of the background from
the processes (13) and (14).}
\end{figure}
%---------------------------Fig13
\newpage
\begin{figure}[htp]
\epsfxsize=15.5cm\epsfysize=12.5cm
\epsffile{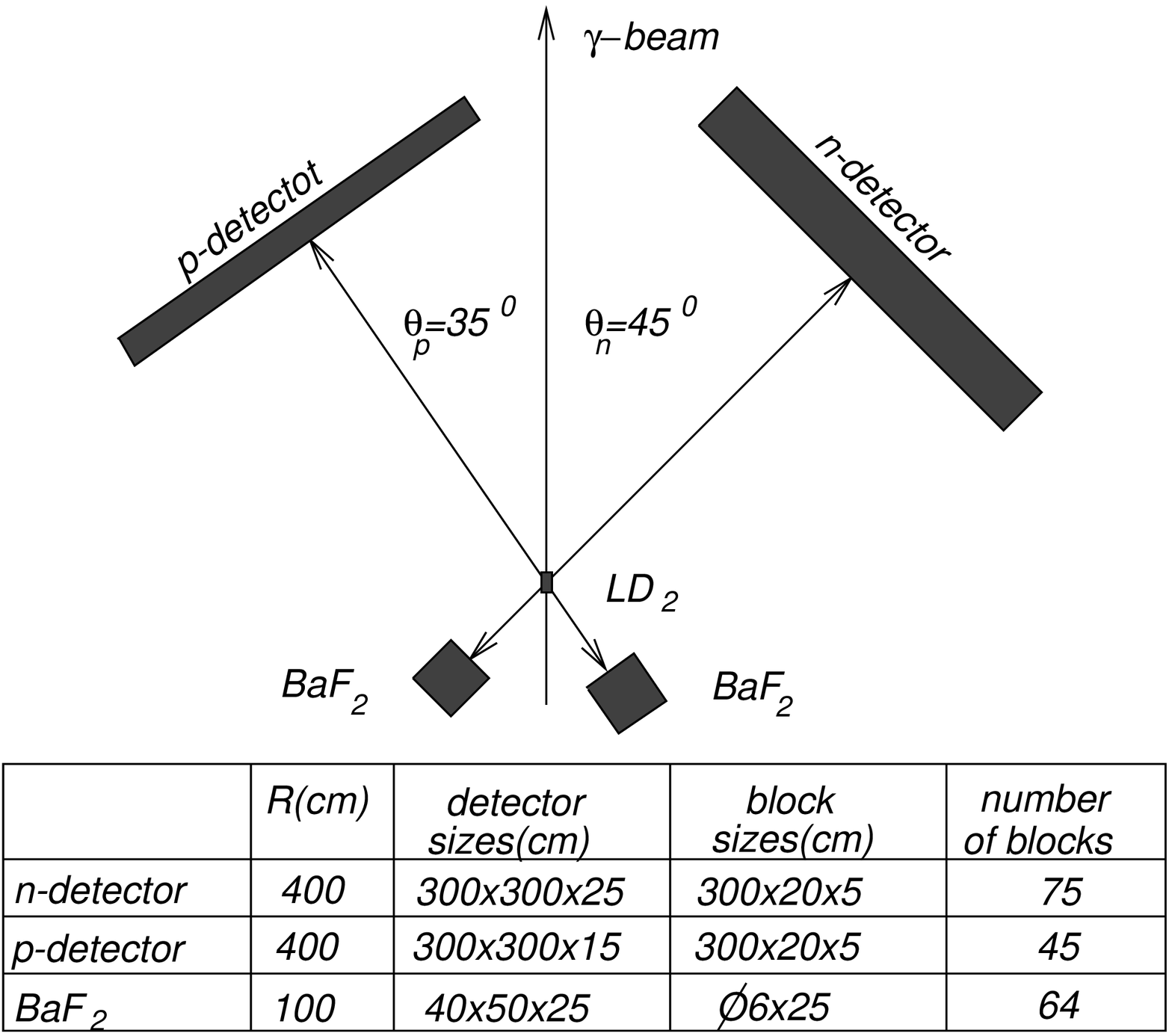}
\vspace*{1.5cm}
\caption{Location of the detectors.}
\end{figure}

\end{document}